\documentclass[11pt,a4paper]{article}
\pdfoutput=1
\usepackage{jcappub}
\usepackage{graphicx}
\usepackage{amssymb}
\usepackage{epsfig}
\usepackage{bm}
\usepackage{dcolumn}
\usepackage{enumerate}
\usepackage{rotating}

\def\eg{{e.g.,~}}
\def\ie{{i.e.,~}}
\def\mnras{{MNRAS}}
\def\apj{{ApJ}}
\def\apjl{{ApJL}}
\def\apjs{{ApJS}}
\def\aap{{AAP}}
\def\prd{{Phys. Rev. D}}
\def\jcap{{J. Cosmol. Astropart. Phys.}}
\def\nat{{Nature}}
\def\araa{{Ann. Rev. Astron. Astrophys.}}

\begin{document}

\title{Axion-like particle imprint in cosmological very-high-energy sources}

\author[1,2,3]{A. Dom\'inguez}
\author[4,5,6]{M.~A. S\'anchez-Conde}
\author[2]{F. Prada}

\affiliation{$^1$Dpto. de FAMN, Universidad de Sevilla, Apdo. Correos 1065, E-41080 Sevilla, Spain}
\affiliation{$^2$Instituto de Astrof\'isica de Andaluc\'ia, CSIC, Apdo. Correos 3004, E-18080 Granada, Spain}
\affiliation{$^3$Dpto. de F\'isica Te\'orica, Universidad Aut\'onoma de Madrid, E-28049 Madrid, Spain; Multidark fellow}
\affiliation{$^4$Instituto de Astrofisica de Canarias, E-38205 La Laguna, Tenerife, Spain}
\affiliation{$^5$Dpto. de Astrof\'isica, Universidad de La Laguna, E-38205 La Laguna, Tenerife, Spain}
\affiliation{$^6$SLAC National Laboratory and Kavli Institute for Particle Astrophysics and Cosmology, 2575 Sand Hill Road, Menlo Park, CA 94025, USA}
\emailAdd{alberto@iaa.es}
\emailAdd{masc@stanford.edu}
\emailAdd{fprada@iaa.es}


\abstract{
Discoveries of very high energy (VHE) photons from distant blazars suggest that, after correction by extragalactic background light (EBL) absorption, there is a flatness or even a turn-up in their spectra at the highest energies that cannot be easily explained by the standard framework. Here, it is shown that a possible solution to this problem is achieved by assuming the existence of axion-like particles (ALPs) with masses $\sim 1$~neV. The ALP scenario is tested making use of observations of the highest redshift blazars known in the VHE energy regime, namely 3C~279, 3C~66A, PKS~1222+216 and PG~1553+113. In all cases, better fits to the observed spectra are found when including ALPs rather than considering EBL only. Interestingly, quite similar critical energies for photon/ALP conversions are also derived, independently of the source considered.
}


\arxivnumber{0000.0000}


\maketitle

\section{Introduction}
Blazars are active galactic nuclei (AGN) characterized by having their jets oriented at small angles towards our line of sight. Their emission covers the entire electromagnetic spectrum from radio up to the most energetic $\gamma$-rays. The synchrotron self Compton (SSC) model provides a successful description for the double-peaked broad spectral energy distribution of low redshift blazars up to several TeV \eg \cite{acciari11,abdo11a}.


Very high energy photons (VHE, 30~GeV--30~TeV) coming from cosmological sources, such as blazars, are attenuated by pair-production with ultraviolet (UV), optical, and infrared (IR) photons in the extragalactic background light (EBL). The EBL is the accumulated radiation due to star formation plus a contribution from AGN emission in galaxies over the entire history of the Universe. As summarized in refs.~\cite{dominguez11a,dominguez11b}, the EBL coming from galaxies seems already well constrained from UV up to mid-IR wavelengths at the minimum intensity level allowed by galaxy counts. This implies the highest transparency to $\gamma$-ray photons from standard physics. Independent EBL data from galaxy counts \cite{madau00,fazio04,keenan10} and different modelings are already in good agreement \cite{franceschini08,dominguez11a,somerville11,gilmore11}, even though direct detection remains elusive due to huge contamination from the zodiacal light \cite{hauser01}. However, a very recent analysis of a data set, which the authors claim to be zodiacal-light free, seems to support the EBL at an intensity level close to galaxy counts as already suggested by other methods \cite{matsuoka11}.  On the other hand, discoveries of VHE photons from high redshift sources by imaging atmospheric Cherenkov telescopes (IACTs) such as MAGIC, VERITAS, and HESS also suggest an EBL intensity level in agreement with galaxy counts and EBL models, \eg \cite{aharonian06b,mazin07,albert08,aleksic11c}. Indeed, at present, the uncertainties in the recovered unattenuated VHE spectra (hereafter, EBL-corrected spectra) of blazars observed with the present generation of IACTs are in general dominated by statistical and systematic effects rather than uncertainties in the EBL modeling \cite{dominguez11a,dominguez11b}.


When referring to differential fluxes, EBL-corrected spectra of some blazars suggest a flatness or even an upturn at the highest observed energies (hereafter, the pile-up problem; see \eg figure~3 in \cite{dominguez11b}), which cannot be easily described by the standard framework described above. The pile-up problem was already noticed some years ago (see \eg refs.~\cite{aharonian02,dwek05}) when a higher intensity EBL was expected. At the time, the most possible solution seemed to be to assume lower EBL intensities (as done \eg in ref.~\cite{aharonian06b}). However, as already discussed, today independent EBL results seem to converge towards a Universe with the highest possible transparency to $\gamma$-ray photons. Therefore, the pile-up problem, if present, can in general no longer be attributed to EBL issues. Modifications to the so-called standard one-zone SSC model have also been proposed in order to solve very hard VHE spectra such as strong relativistic shocks \cite{stecker07}, selection bias effects \cite{persic08}, internal absorption \cite{aharonian08}, production in extended jets \cite{boettcher08}, but there is still little work focused in solving the pile-up problem \eg \cite{aharonian02}. Indeed, there is not any general agreement on any definitive solution yet. The postulated existence of axion-like particles (ALPs) arises as an alternative solution to the problem. In fact, the main goal of this paper is to show explicitly that the theoretical framework developed in ref.~\cite{sanchez-conde09} when applied to measured VHE spectra of distant blazars is a possible solution that may deserve further attention.

\begin{sidewaystable}[!h]
\centering
\begin{tabular}{l|c|c|c|c|c|}
\cline{2-6}
& \multicolumn{5}{c|}{Source name}\\
\hline
\multicolumn{1}{|c|}{Parameters} & 3C~279 (2006) & 3C~279 (2007) & 3C~66A & PKS~1222+216 & PG~1553+113\\
\hline
\hline
\multicolumn{1}{|l|}{$R$ [pc]} & 0.2 & 0.3 & 0.4 & 0.003 & 0.03\\
\multicolumn{1}{|l|}{$B$ [G]} & 0.15 & 0.10 & 0.01 & 0.28 & 0.5\\
\multicolumn{1}{|l|}{$n_{e}$ [cm$^{-3}$]} & 19992 & 2.22 & 0.42 & 668446 & 1.97\\
\multicolumn{1}{|l|}{$L$ [pc]} & 0.02 & 0.03 & 0.04 & 3$\times 10^{-4}$ & 0.003\\
\hline
\multicolumn{1}{|l|}{$z$} & 0.536 & 0.536 & 0.444 & 0.432 & 0.4\\
\multicolumn{1}{|l|}{$B_{IGM}$ [nG]} & 0.1 \& 1 & 0.1 \& 1 & 0.1 \& 1 & 0.1 \& 1 & 0.1 \& 1\\
\multicolumn{1}{|l|}{$n_{IGM}$ [cm$^{-3}$]} & 10$^{-7}$ & 10$^{-7}$ &  10$^{-7}$ & 10$^{-7}$ & 10$^{-7}$\\
\multicolumn{1}{|l|}{$L_{IGM}$ [Mpc]} & 1 & 1 & 1 & 1 & 1\\
\hline
\multicolumn{1}{|l|}{$M$ [GeV]} & 1.14 $\times$ 10$^{10}$ & 1.14 $\times$ 10$^{10}$ & 1.14 $\times$ 10$^{10}$ & 1.14$\times$ 10$^{10}$ & 1.14$\times$ 10$^{10}$\\
\multicolumn{1}{|l|}{$m_{a}$ [neV]} & $0.13\le m_{a} \le 1.45$ & $0.13\le m_{a} \le 1.45$ & $0.13\le m_{a} \le 1.45$ & $0.13\le m_{a} \le 1.45$ & $0.13\le m_{a} \le 1.45$\\
\hline
\end{tabular}

\caption{\small{Parameters used to calculate the total photon/ALP conversion in both the source (for the four AGNs considered, 3C~279, 3C~66A, PKS~1222+216 and PG~1553+113) and in the IGM. $R$ is the size of the region where $B$ is confined within the source, calculated as discussed in ref.~\cite{sanchez-conde09}, while $L$ is the length of the coherent domains inside $R$. Note that we chose $R=10 \times L$ in all cases. $n_{e}$ is the electron density in the source. As for the IGM, its density $n_{IGM}$ was taken from ref.~\cite{peebles93}. See text for details.}}
\label{tab1}
\end{sidewaystable}

\section{Theoretical framework}
Axions are predicted by the Peccei-Quinn mechanism, which is currently the most compelling explanation to solve the CP problem in quantum chromodynamics \cite{peccei77}. They might constitute a portion or the totality of the non-baryonic cold dark matter content in the Universe \cite{steffen09}. One may also consider the existence of ALPs, which are particles with the same properties as axions, but with mass and coupling constant not related to each other \cite{masso95}. There is an additional property of ALPs that could have important implications for their detectability as well as for $\gamma$-ray astronomy, \ie they can convert into photons and vice-versa in the presence of an electric or magnetic field \cite{dicus78,sikivie83}. This argument was first investigated in the optical band by ref.~\cite{csaki02}, where authors proposed the existence of axions to be the cause of the observed supernova Ia dimming (this proposal is ruled out at present; see \eg ref.~\cite{mirizzi08} and references therein). Later, the photon/ALP conversion was applied for the same authors at higher energies in ref.~\cite{csaki03}. The complementary view of photon/ALP oscillations both in the source and in the Galactic magnetic fields was first put forward in ref.~\cite{simet08}. It has been also proposed that ALPs might significantly affect $\gamma$-ray AGN observations \cite{hooper07,deangelis07,deangelis09,deangelis11}. Ref.~\cite{sanchez-conde09} was the first work in offering a complete formalism that includes under the same consistent framework the photon/ALP mixing expected to occur in both the $\gamma$-ray sources and in the intergalactic medium (IGM). There, prospects for ALP detection were also discussed. One possible approach lies in detecting intensity boost in the higher energy bins of cosmological VHE observations by IACTs as a consequence of an efficient photon/ALP conversion. Other approaches might be considered as well, such as detecting abrupt changes in intensity at a given critical energy, $E_{crit}$, defined in GeV as $E_{crit}=m_{\mu eV}^{2}M_{11}/0.4B_{G}$, which should be the same for all the sources \cite{hooper07,deangelis08}. In this equation, $M_{11}$ is the photon/ALP coupling constant divided by $10^{11}$~GeV, $B_{G}$ the magnetic field in units of G, and $m_{\mu eV}$  the effective ALP mass in $\mu$eV, defined as $m_{\mu eV}^{2} \equiv |m_{a}^{2}-\omega_{pl}^{2}|$, where $m_a$ is the ALP mass and $\omega_{pl}=0.37 \times 10^{-4} \sqrt{n_e}$ the plasma frequency in $\mu$eV, $n_e$ being the electron density in units of cm$^{-3}$. These alternative approaches will be followed elsewhere.

\begin{figure*}[ht]
\includegraphics[trim=4cm 0cm 5cm 0cm, width=\textwidth]{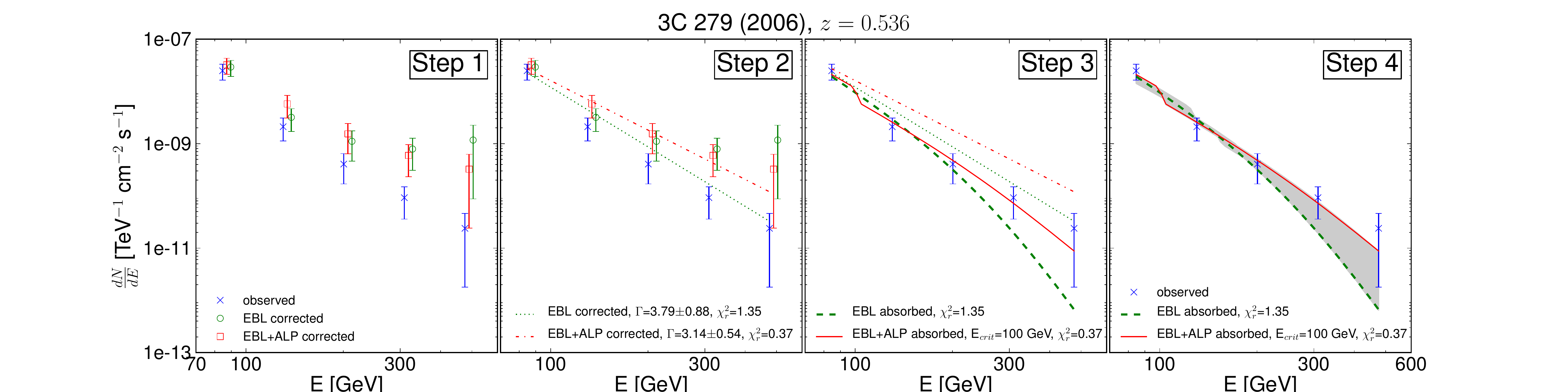}
\caption{\small{The effect of ALPs on the spectrum of 3C~279 obtained from 2006 MAGIC data \cite{albert08}. Data points corrected for EBL only and for EBL plus ALP in Steps~1 and 2 are slightly shifted for clarity. See the text for a detailed explanation of every step.}}
\label{fig:3C2792006}
\end{figure*}

\section{Application to distant known blazars}
Our methodology for studying any possible hint of a signature introduced by ALP in the observed VHE blazar spectra is based on several assumptions: (i) the intrinsic spectra emitted from the sources are well described by simple power laws, which is a good approximation for the relatively small energy ranges considered in the blazars studied here; (ii) $M_{11}$ is an optimistic value, so that we can get maximum photon/ALP conversions, but still within present experimental limits, \ie those imposed by the CAST experiment \cite{andriamonje07}\footnote{We note that significantly higher $M$ values will probably lead to a negligible effect due to photon/ALP conversions in the sources. Yet, the effect coming from conversions in the intergalactic magnetic field is not negligible, but would be substantially lower than the one shown in this paper. As a consequence, ALPs might not probably work fine to solve the pile-up problem in many sources.}; (iii) the value of the critical energy corresponding to the intergalactic mixing (hereafter, $E_{crit}$) lies in the energy range measurable by IACTs, or below; (iv) the EBL is well described by the model described in ref.~\cite{dominguez11a}. 

The effect of ALP on $\gamma$-rays being emitted by blazars is modeled as follows: (i) the intrinsic emission from every source is obtained from the best SSC model fit to the multiwavelength observations for each source; (ii) through the IGM we adopt two different values of the intergalactic magnetic field $B_{IGM}=0.1$~nG and $B_{IGM}=1$~nG given its large uncertainties at present \eg \cite{abraham08}; (iii) once $E_{crit}$ and $B_{IGM}$ are fixed, we then calculate the ALP mass $m_{a}$ using the relation between these quantities described in the above formula. In this study, we scan $E_{crit}$ within the range 50--625~GeV in steps of 25~GeV, which corresponds to $m_{a}\sim 1$~neV. It is important to note that any $E_{crit}$ below 50~GeV will lead, in the energy range under consideration, to identical results than those obtained for 50~GeV. The reason is the strong dependence of the ALP effect on the EBL intensity. As the role of the EBL is negligible below $\sim 50$~GeV, photon/ALP mixings up to this energy will produce identical features in the spectra. Relevant source, IGM, and ALP parameters are listed in Table~\ref{tab1}.


We now carefully describe our methodology when applied to 3C~279, which is later repeated for other potentially interesting blazars, in particular 3C~66A, PKS~1222+216 and PG~1553+113. These four objects were selected according to their high redshift, where the effect of photon/ALP conversion in observed spectra is expected to be larger at the highest measured energies \cite{sanchez-conde09}. Indeed, this sample actually contains the four highest-redshift blazars observed so far and for which there is published spectra available.

\begin{figure*}[t,b]
\centering
\includegraphics[trim=1cm 0cm 1cm 0cm, width=7.6cm]{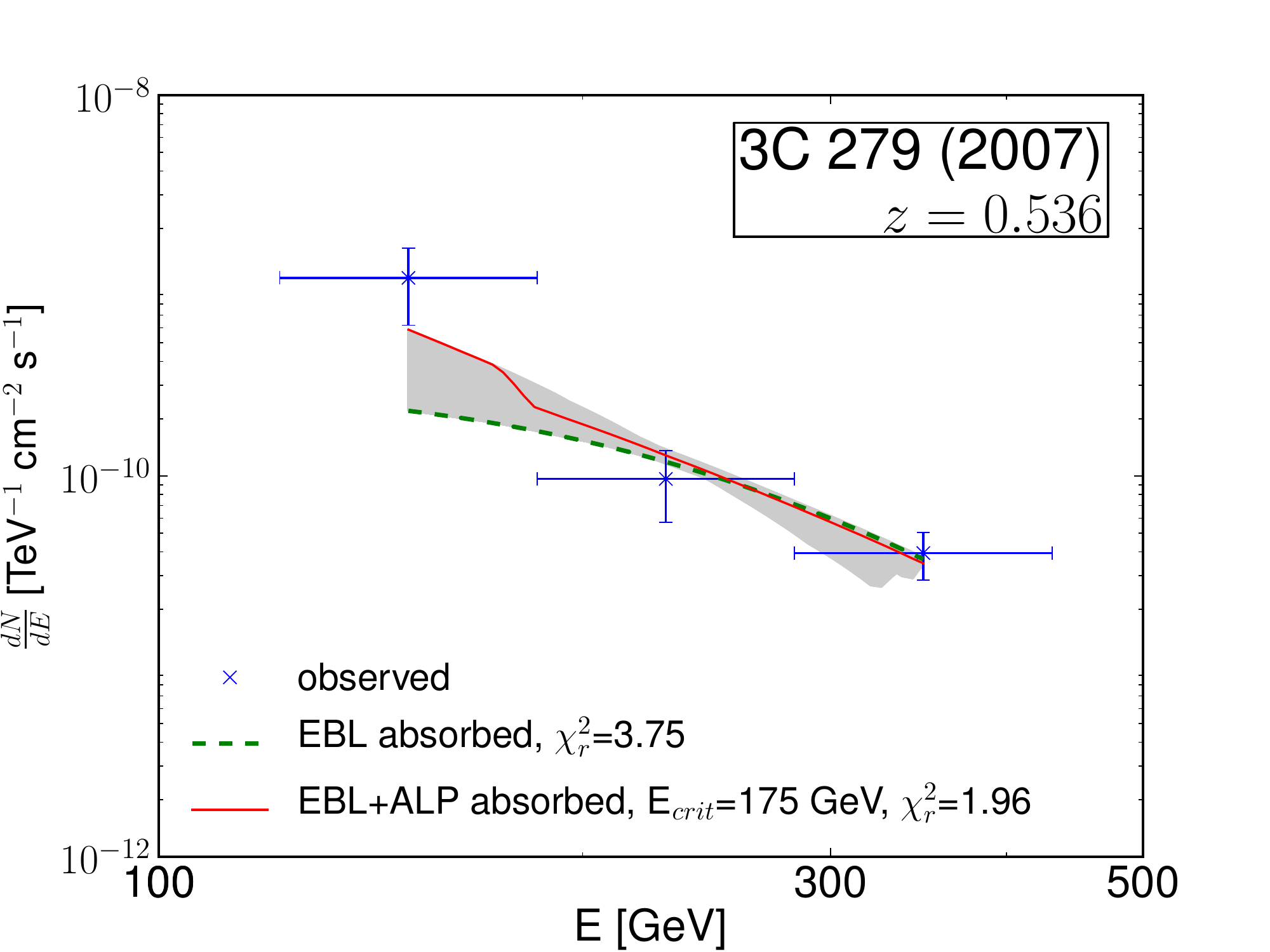}
\includegraphics[trim=1cm 0cm 1cm 0cm, width=7.6cm]{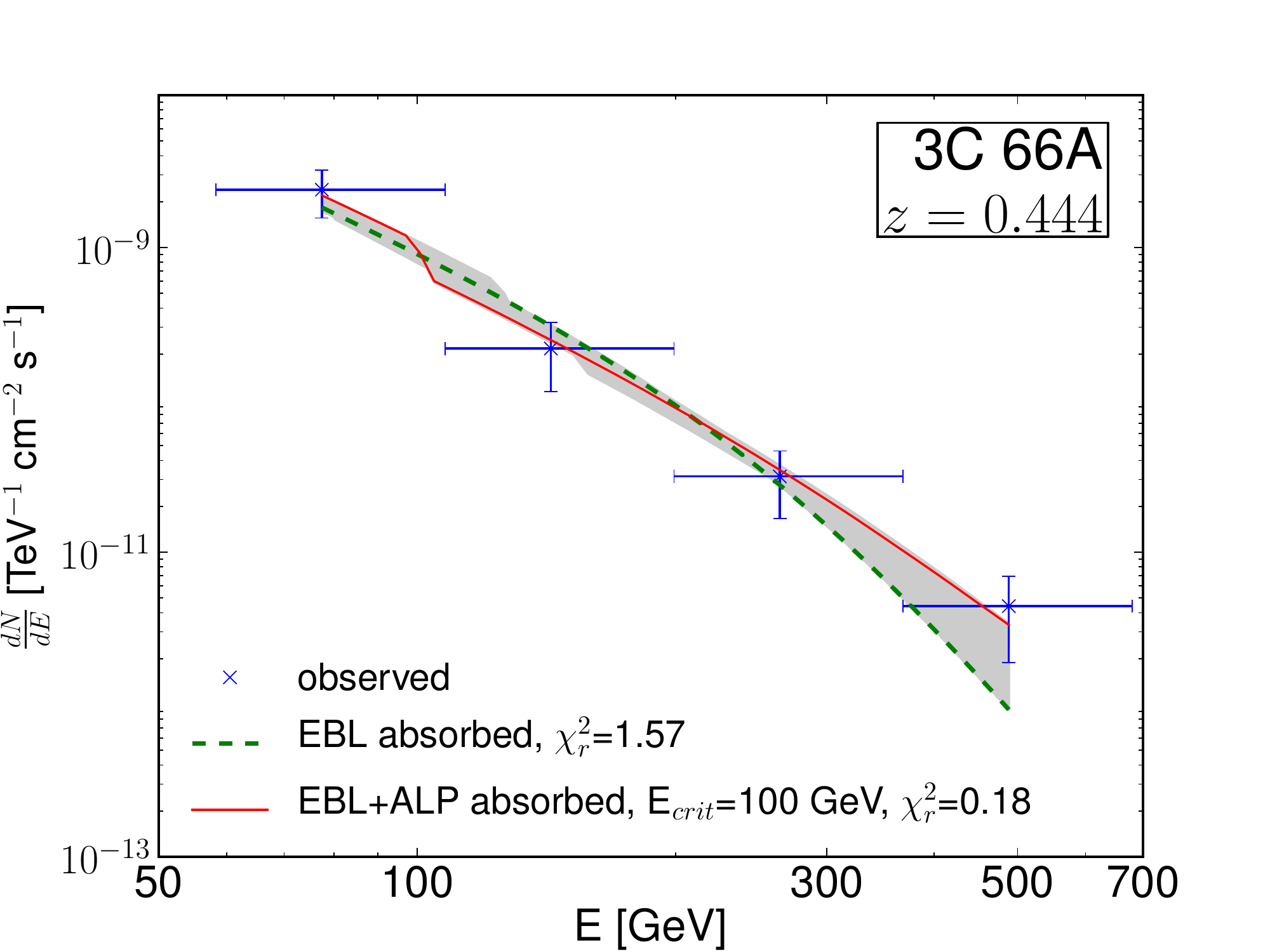}
\includegraphics[trim=1cm 0cm 1cm 0cm, width=7.6cm]{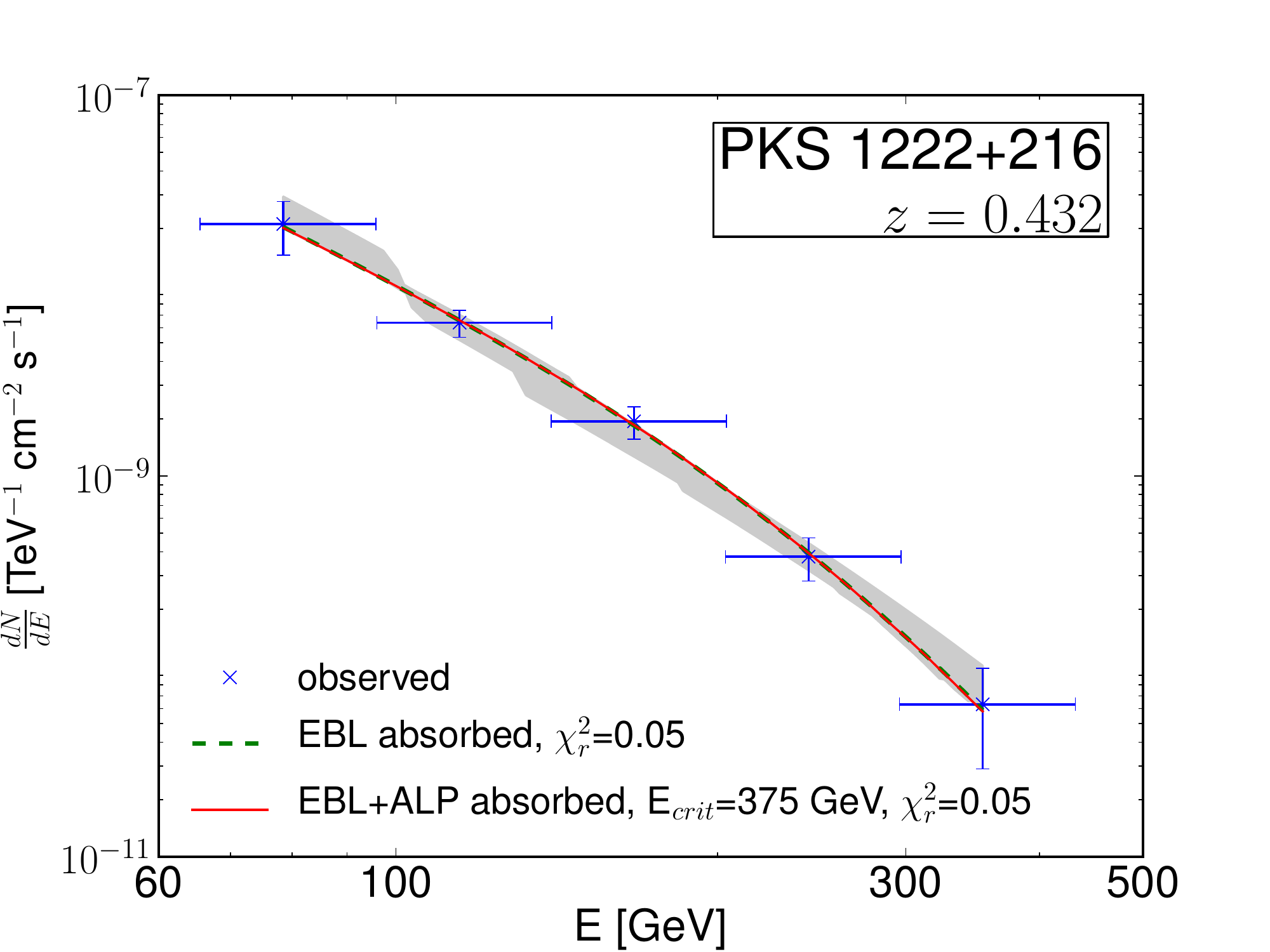}
\includegraphics[trim=1cm 0cm 1cm 0cm, width=7.6cm]{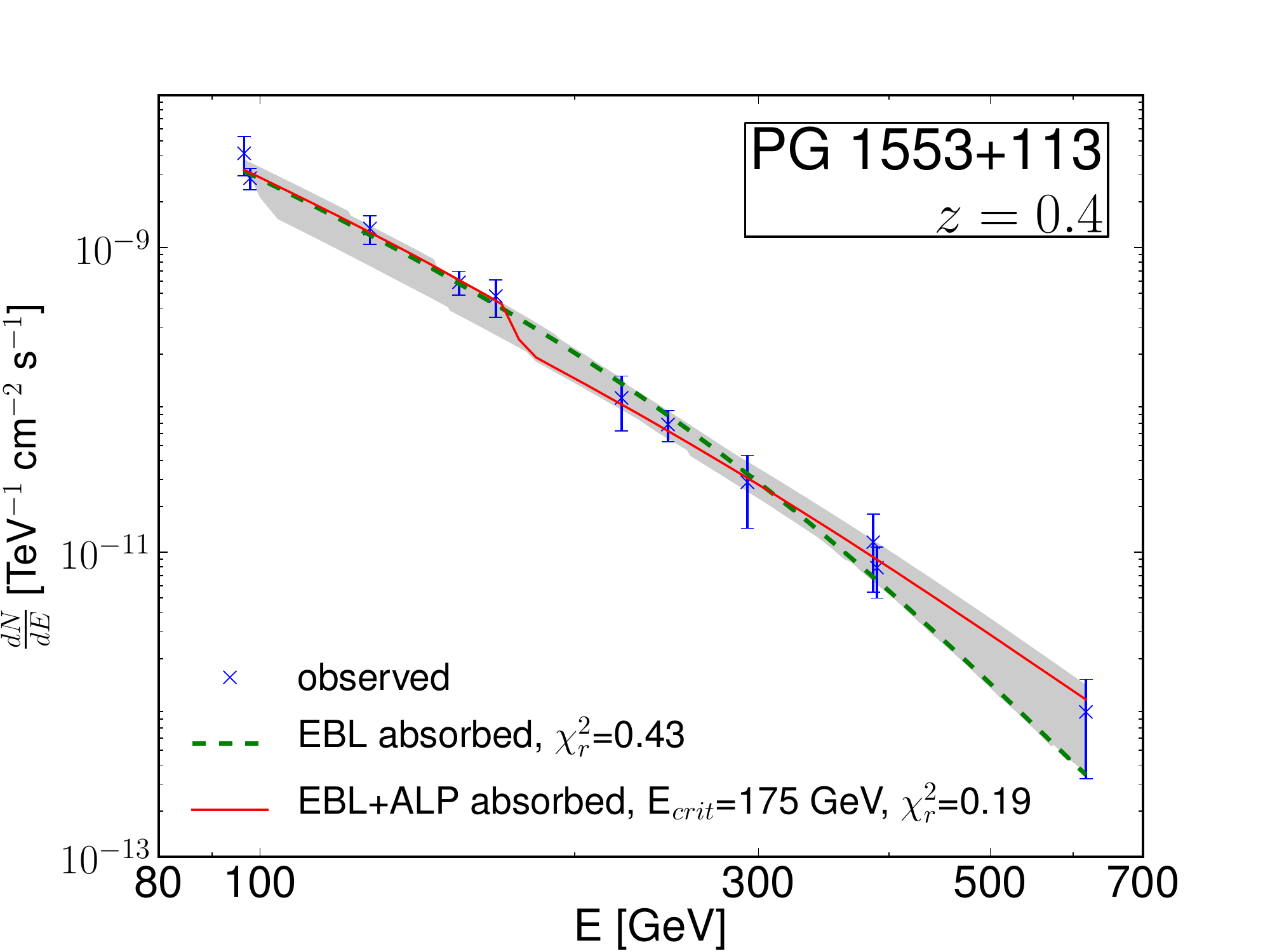}
\caption{\small{The effect of ALPs on high redshift VHE observations. \emph{Upper-left panel:} 3C~279 observation by MAGIC in the 2007 observational campaign \cite{aleksic11b}. \emph{Upper-right panel:} 3C~66A observation by MAGIC \cite{aleksic11a}. \emph{Lower-left panel:} PKS~1222+216 discovery by MAGIC \cite{aleksic11c}. \emph{Lower-right panel:} PG~1553+113 extensive campaign from 2005-2009 by MAGIC \cite{aleksic11d}. As additional case, the 3C~279 spectrum observed by MAGIC in 2006 was already shown in figure~\ref{fig:3C2792006} (Step~4).}}
\label{fig:blazars}
\end{figure*}

\subsection{3C~279}
We show in figure~\ref{fig:3C2792006} and figure~\ref{fig:blazars} (upper-left panel) VHE data from 3C~279 ($z=0.536$). These spectra were measured by the MAGIC collaboration at two different epochs \cite{albert08,aleksic11b} when this flat spectrum radio quasar (FSRQ) was in different flaring states. To model the photon/ALP mixing within the source, we use the SSC parameters given for both epochs in ref.~\cite{aleksic11b} (see Table~\ref{tab1}). Different steps are followed in order to obtain our final results. These steps are only explicitly shown in figure~\ref{fig:3C2792006} for the 2006 data, while only the last step is plotted in figure~\ref{fig:blazars} for the other spectra used in this study. Step 1) The observed VHE spectrum (blue crosses) is corrected by the effect from the EBL only (green circles), and by the EBL plus ALPs (red squares). The case with $E_{crit}=100$~GeV and $B_{IGM}=0.1$~nG is plotted in figure~\ref{fig:3C2792006} as an example. The pile-up problem is evident here for the EBL-corrected spectrum (green circles). Step 2) Both corrected (or de-absorbed) spectra are fitted by simple power laws. The spectral indexes and the $\chi_{r}^{2}=\chi^{2}/n$ of the fits are shown in figure~\ref{fig:3C2792006} (with $n$ degrees of freedom). Step 3) The best-fit power laws obtained from Step 2 are then absorbed again by the EBL (dashed green line), and the EBL plus ALP (solid red line). Although small, the drop in flux at $E_{crit}$ (in this particular case 100~GeV) is still noticeable. This step is taken in order to check how well the observed spectrum is recovered in the case where we account for only the EBL effect and when including EBL plus ALPs. The reduced $\chi^{2}$ labeled in the figure corresponds to the corrected best-fit power law given in Step~2. Step 4) Steps 1 to 3 are then repeated by varying $E_{crit}$ in 25~GeV steps within the energy range in consideration and using the two different values of $B_{IGM}$ considered here (see Table~\ref{tab1}). The best fit among all the cases is plotted in figure~\ref{fig:3C2792006} (right panel) as a solid red line. For this particular source, the best fit is obtained for $E_{crit}=100$~GeV. Note that this is indeed a better fit to the data than when only considering the EBL attenuation (dashed green line). This is quantitatively shown in Table~\ref{tab2}, where we provide, for all sources, the $\chi_{r}^{2}$ of the fits for five benchmark $E_{crit}$ in the spectra. Finally, the shadow region in every figure represents the set of fits obtained when varying $E_{crit}$ within the whole energy interval, in 25 GeV steps as mentioned above. Therefore, this shadow region might be taken as the uncertainty region that arises due to our lack of knowledge of the exact $E_{crit}$. Note that $E_{crit}$ does not enter as an additional parameter in the fits; indeed, both kind of fits (\ie with and without ALPs) do actually have the same degrees of freedom. The uncertainty region simply reflects a lack of knowledge on the underlying physics (real ALP parameters and $B_{IGM}$), each particular fit in this region being completely independent of the others (meaning that $E_{crit}$ does not act as a linking parameter between them).

\begin{table*}
\centering
\begin{tabular}{lc|c|c|c|c|c|}
\cline{3-7}
& & \multicolumn{5}{c|}{$E_{crit}$ [GeV]}\\
\hline
\multicolumn{1}{|c}{Source name} & \multicolumn{1}{|c|}{no ALPs} & 100 &
200 & 300 & 400 & 500 \\
\hline
\hline
\multicolumn{1}{|c}{3C~279 (2006)} & \multicolumn{1}{|c|}{1.35} & {\bf
0.37/0.42} & 1.11/1.07 & 1.10/1.07 & 1.28/1.28 & 1.36/1.36 \\
\multicolumn{1}{|c}{3C~279 (2007)} & \multicolumn{1}{|c|}{3.75} &
3.04/3.10 & {\bf1.96/2.19} & 2.23/2.07 & 3.74/3.74 & 3.74/3.74 \\
\multicolumn{1}{|c}{3C~66A} & \multicolumn{1}{|c|}{1.57} & {\bf 0.18/0.24}
& 0.89/0.86 & 0.88/0.87 & 0.88/0.87 & 1.34/1.34\\
\multicolumn{1}{|c}{PKS~1222+216} & \multicolumn{1}{|c|}{0.05} & 1.47/1.35
& 0.18/0.18 & 0.09/0.14 & {\bf 0.05/0.05} & 0.05/0.05\\
\multicolumn{1}{|c}{PG~1553+113} & \multicolumn{1}{|c|}{0.43} & 1.39/1.28 & {\bf 0.19/0.18} & 0.26/0.23 & 0.32/0.32 & 0.32/0.32\\
\hline
\end{tabular}
\caption{\small{The $\chi^{2}_{r}$ values of the power-law fits to the de-absorbed spectra by the EBL effect only, and EBL plus ALPs for five benchmark values of the intergalactic $E_{crit}$. Note that two $\chi^{2}_{r}$ are given for every $E_{crit}$, corresponding to the case with $B_{IGM}=0.1$~nG and $B_{IGM}=1$~nG, respectively. In bold, the best $\chi_{r}^{2}$ for every source. Note that for PKS~1222+216 and PG~1553+113 cases, all $E_{crit}$ have roughly the same low $\chi_{r}^{2}$, except $E_{crit}$=100~GeV.}}
\label{tab2}
\end{table*}

\subsection{3C~66A}
This source at $z=0.444$ was observed in a flaring state by VERITAS \cite{acciari09} and more recently by MAGIC \cite{aleksic11a}. Both spectra are roughly compatible within uncertainties. Note that the redshift considered for this object is calculated using just one emission line and is thus not very secure, as discussed in ref.~\cite{bramel05}. In this study, only the MAGIC spectrum is used due to its larger energy coverage, which better exposes the pile-up problem. We model the source using the parameters given in Table~\ref{tab1}, which were extracted from ref.~\cite{abdo11b} and were computed by the authors using an SSC model. Figure~\ref{fig:blazars} shows the 3C~66A spectrum as measured by MAGIC (upper-right panel). The best fit is shown following the same methodology as in figure~\ref{fig:3C2792006} for the case of 3C~279. We only show the results corresponding to Step~4. Again, considering the existence of ALP provides a better fit.



\subsection{PKS~1222+216}
This FSRQ located at $z=0.432$ was discovered in the VHE regime in a flaring state by the MAGIC collaboration \cite{aleksic11c}. We model the source using the SSC parameters given in \cite{tavecchio11} for the so-called \emph{all blob} emission model. In any case, we stress that the effect of the photon/ALP conversion within the source is estimated to be less than 10\% of the total. In this particular case, the mixing within the source is negligible mainly due to the small value of the region where the magnetic field is confined ($R$, see Table~\ref{tab1}) and therefore only intergalactic mixing is present. The results obtained for this blazar are shown in figure~\ref{fig:blazars} (lower-left panel). In this case, both fit (EBL only and EBL plus ALPs) good. This is important, as it represents a good test of our ALP formalism: in those cases where the data are well understood and reproduced by conventional physics (regarding EBL and source modeling), ALP should not introduce any additional effect and should therefore be irrelevant if included.

\subsection{PG~1553+113}
This blazar was first detected in the VHE regime by the HESS collaboration in 2005 \cite{aharonian06a}. Since then, it has been extensively observed by the MAGIC collaboration, whose results are detailed in ref.~\cite{aleksic11d}. In our study, the combined VHE data from this five-year campaign by MAGIC are used. The SSC modeling is also taken from ref.~\cite{aleksic11d}. On the other hand, the PG~1553+116 redshift has remained elusive (see \eg ref.~\cite{prandini10}), even though there are upper limits ($z<0.614$ at $1\sigma$) coming from the expected EBL attenuation \cite{prandini10}. Recently, \cite{danforth10} estimated its redshift at $z=0.4-0.45$ from spectral Ly-$\alpha$ signatures. For the redshift of the source we assume $z=0.4$, warning the reader about its uncertainty. The result from applying our ALP formalism to this blazar is shown in figure~\ref{fig:blazars} (lower-right panel). In this case, as in the PKS~1222+216 case, both the EBL-only and EBL-plus-ALPs curves provide a good fit to the observed data. Again, this tests that our ALP formalism does not introduce any additional effect in those cases where conventional physics by itself represents a good fit to the data.

\section{Discussion}
We have shown a possible alternative explanation to the pile-up problem observed at the highest energies in the EBL-corrected VHE spectra of the most distant blazars. This feature indeed sets a challenge to our current understanding of the emission mechanisms in blazars and/or the propagation of VHE photons through the IGM. While spectra corrected only for EBL absorption hardly describe some observations, the spectra corrected taking into account the existence of ALPs (with $m_{a}\sim 1$~neV) provides a better fit to the data for all the objects discussed here, i.e. 3C~279, 3C~66A, PKS~1222+216 and PG~1553+113, independently of source modeling. Indeed, the pile-up problem is completely alleviated when including ALPs in the first two cases. On the other hand, in the case of PKS~1222+216 and PG~1553+113, the pile-up feature is not present and therefore ALPs are not needed; however ALPs provide a good fit as well, which actually represents a good check of our formalism (as ALPs would be always present if they exist). It is important to stress that these better fits to the data are not achieved by introducing any additional parameter in the fitting procedure. Indeed, the number of parameters involved in all the fits remains the same and equal to the degrees of freedom allowed by a simple power law. In this sense, we note that ALP parameters (such as $E_{crit}$) play a role which is just comparable to that played by those parameters describing the EBL model/physics. Both EBL-only and EBL+ALP curves shown in the plots correspond indeed to the outputs of different underlaying physical frameworks, each of one with their own parameters; both sets of parameters do not enter, in any case, in the \emph{posterior} power-law fits to the data.

It is also noticeable that those blazars that profit from the existence of ALPs roughly agree on the $E_{crit}$ that better fits the data ($\sim100-200$~GeV). Indeed, for each particular case, the most appropriate $E_{crit}$ lies near the lowest energy data point available. We note that every $E_{crit}$ adopted was obtained by assuming two different $B_{IGM}$ values; this fact points towards a degeneracy of our results for $B_{IGM}$. However, in practice, this degeneracy actually leads to very similar results (see Table~\ref{tab2}). A way to break this degeneracy would be by observing AGN spectra to the highest possible energies, where different $B$ values should lead to substantially different intensity boosts.

We do not claim here to have shown that ALPs indeed exist. Other possible solutions to the pile-up problem within standard physics may exist such as large statistical and systematic uncertainties in the measurements, extensions of the SSC model \cite{aharonian02,stecker07,persic08,aharonian08,boettcher08}, or $\gamma$-ray photons from secondary interactions \cite{aharonian02,essey11}. Better source statistics, an extension of the analysis to lower-redshift sources, and a more detailed study are needed, but we aimed here to show explicitly in real VHE spectra of distant sources that the ALP possibility does deserve further study. We can envision multiple consequences of some of the ingredients with a prominent role in the ALP formalism discussed here. For instance, note that in the case that ALPs exist with the properties adopted throughout this paper, the EBL upper limits derived from observation of VHE spectra of blazars \cite{aharonian06b,mazin07,albert08} should be revised.

\acknowledgments
The authors thank Fabrizio Tavecchio and Luis C. Reyes for their help in calculating SSC parameters, and the anonymous referee for helpful suggestions. We thank the support of the Spanish MICINN's Consolider-Ingenio 2010 Programme under grant MultiDark CSD2009-00064. A.D.'s work has also been supported by the Spanish MEC and the European FEDER under projects FIS2008-04189 and CPAN-Ingenio (CSD2007-00042), and by the Junta de Andaluc\'ia (P07-FQM-02894). A.D. is grateful for the hospitality of the Instituto de Astrof\'isica de Canarias during his visit.


\label{lastpage}
\end{document}